\documentclass[seceq]{ptptex}
\usepackage{graphicx,color,amsmath,amssymb}

\usepackage{cancel}

\usepackage{geometry}
\newcommand{\Slash}[1]{\ooalign{\hfil/\hfil\crcr$#1$}}
\newcommand{\pa}{\partial}
\newcommand{\nn}{\nonumber}

\newcommand{\Psibar}{{\bar \Psi}}
\newcommand{\ep}{\epsilon}
\newcommand{\varep}{\varepsilon}
\newcommand{\bref}[1]{(\ref{#1})}

\newcommand{\ds}{{\displaystyle}}

\begin{document}
\title{Realization of chiral symmetry in the ERG}
\author{Yoshio Echigo and Yuji Igarashi${}^a$}
\maketitle
\vspace{-0.3cm}
\begin{center}
Graduate School of Science and Technology, Niigata University, Niigata, 950-2181, JAPAN
${}^a$ Faculty of Education, Niigata University, Niigata, 950-2181, JAPAN\\
\end{center}
\vspace{0.8cm}
\begin{abstract}
We discuss within the framework of the ERG how chiral symmetry is
realized in a linear $\sigma$ model.  A generalized Ginsparg-Wilson
relation is obtained from the Ward-Takahashi identities for the Wilson
action assumed to be bilinear in the Dirac fields. We construct a family
of its non-perturbative solutions. 
The family generates the most general solutions to the Ward-Takahashi
 identities. Some special solutions are discussed. 
For each solution in this family, chiral symmetry is realized
in such a way that a change in the Wilson action
under non-linear symmetry
transformation is canceled with a change in the functional measure.  We
discuss that the family of solutions reduces via a field redefinition to
a family of the Wilson actions with some composite object of the
scalar fields which has a simple transformation property. For this
family, chiral symmetry is linearly realized with a continuum analog of
the operator extension of $\gamma_5$ used on the lattice. We also
 show that there exist some appropriate Dirac fields which obey 
the standard chiral transformations with $\gamma_5$ in contrast to the
 lattice case. Their Yukawa interactions with scalars, however,
 becomes non-linear.

\end{abstract}


\section{Introduction}
\setcounter{page}{1}
\setcounter{footnote}{0}
The discovery of chiral symmetry on the lattice\cite{GW}\tocite{Luesher}
is considered as a new realization of symmetry which is incompatible
with a given regularization used in a perturbative and/or
non-perturbative approach to field theory.  The crucial issue in this
realization lies in an algebraic non-linear constraint on the Dirac
operators, the Ginsparg-Wilson (GW) relation\cite{GW}. It makes possible
to define symmetry transformations\cite{Luesher} which are not simply
specified by the standard $\gamma_5$ matrix, but its non-trivial
extension that depends on the Dirac operator as well as the lattice
spacing, avoiding the no-go theorem\cite{Nielsen-Ninomiya}.  It has been
also shown that this formulation of chiral symmetry correctly gives the
index theorem related to the chiral anomaly\cite{Hasenfratz}\cite{Luesher}.

The GW relation is essentially derived for free-field theories. Gauge
field can be easily included as an external field in the relation. It is
non-trivial to discuss a generalized GW relation in the presence of
interactions of fermions with other fields which undergo chiral
transformations.

In this paper we discuss this issue in the context
of the exact renormalization group (ERG)\cite{Wilson}\tocite{Wetterich} 
 approach\footnote{Recent reviews on the ERG are 
refs.~\citen{Litim}~--~\citen{Igarashi-Itoh-Sonoda1}.} to continuum theory.
The original GW relation was derived for a macroscopic action of the
Dirac fields on a coarse lattice obtained via the block-spin
transformations from a microscopic action on a fine lattice. In parallel
with this, the Wilson action in the ERG 
defined at IR scale $\Lambda$ is obtained by blocking of an
action given at a UV scale $\Lambda_0$. We consider a symmetry 
assumed to be realized in the standard form at $\Lambda_0$. It suffers from a
deformation when our reference scale $\Lambda$ is going down from
$\Lambda_0$ as we take a symmetry-breaking blocking. Even though the
symmetry cannot remain in the standard form, it is realized in a
non-trivial way. In general, the deformation of symmetry and its properties are
described by the Ward-Takahashi (WT) identities for the Wilson action,
and by their extended version, the Quantum Master Equation (QME) in the
Batalin-Vilkovisky\cite{BV} antifield formalism\footnote{See, e.g., 
refs.~\citen{Becchi}~\citen{Igarashi-Itoh-So1} for the generic idea of
formulation of exact (gauge) symmetry in ERG, and 
refs.~\citen{Sonoda}~--~\citen{Igarashi-Itoh-Sonoda3} for some concrete
discussions for QED and Yang-Mills theory.}.  
 
If we confine ourselves to those for the Wilson action that is bilinear
in the Dirac fields, the WT identities for the fermionic sector lead to
a relation which is quadratic in the Dirac operator.  For free-field
theories, we have discussed that the WT identities lead to the standard
GW relation for chiral symmetry and that for SU(2) global
symmetry\cite{Igarashi-Itoh-So2}.

In the presence of interactions between the Dirac fields and some other
fields, the quadratic relation in the Dirac operator may be called as a
generalized GW relation which contains the fields interacting with the
Dirac fields. This generally happens, not restricted to chiral
symmetry. Consideration of such a GW relation in chiral symmetry,
therefore, may give some important insights into realization of other
symmetries including that of gauge symmetry.

We wish to discuss in this paper chiral symmetry in a linear $\sigma$
model.  We take symmetry-breaking blocking with a mass term for the
Dirac fields to discuss non-trivial realization. Derivation of the WT
identities for a generic linear symmetry has been discussed in
ref.~\citen{Igarashi-Itoh-Sonoda2}, and also in
ref.~\citen{Bergner-Bruckmann-Pawlowski}.  In the functional
integral method we use, the Wilson action for the IR fields 
is formally defined by an integration of the blocked UV action to
construct the WT identities for the IR action. 
Once we obtain the identities, we try
to discuss the most general solutions to them in a purely algebraic
way. The UV action is only used as
a boundary condition for the solutions, and is not used to be integrated
out.  Therefore, we are
not confined ourselves to those solutions obtained from explicit
integration of the UV fields. 
Our main task is to solve the
WT identities for the Wilson action for the $\sigma$ model which is
assumed to be bilinear in the Dirac fields.  Although this is a
truncation in the fermionic sector, the WT identities we obtain are
exact if we neglect a loop contribution arising from four-fermi
interactions. The resulting identity for the Dirac fields leads to a
generalized GW relation for the Dirac operator which has a non-linear
dependence on the scalar fields.  We obtain the most general 
non-perturbative solutions to the GW relation. 
They include as a special case 
the continuum counter part of the solutions obtained on
the lattice in ref.~\citen{Igarashi-So-Ukita}, where the scalar fields are
introduced as auxiliary fields in the Wilson action. There is also some
other choice which corresponds to the solution obtained by explicit
integration over the UV Dirac fields. It corresponds to a perfect action
if no blocking for the scalar fields is performed.

It should be stressed that the chiral transformations for the Dirac
fields depend on the Dirac operator, and become non-linear. They induce
a Jacobian factor that must be canceled with a counter action of the scalar   
 fields. Thus, the Wilson action constructed from a solution to the GW
 relation contains the counter action which is not chiral
 invariant. Chiral symmetry is not realized as invariance of the Wilson
 action, and inclusion of the contribution from change in the functional
 measure is crucial in this realization.

On the lattice, an operator extension of $\gamma_5$ matrix denoted as
$\hat\gamma_5$ is extensively used. It depends on the free Dirac
operator and the lattice spacing.  In the ERG approach to continuum
theory considered in this paper, there exists a field redefinition which
reduces our family of the solutions to a family of the Wilson actions
for which the chiral transformation on the Dirac fields is expressed in
terms of $\hat\gamma_5$.  Each Wilson action consists of some composite
object of the scalar fields which has a simple transformation
properties.  Using new fermionic variables, chiral transformation is
linearly realized, and symmetry is simply realized as invariance of the
Wilson action. We also show that appropriate field redefinition gives
the Dirac fields which obey the standard chiral transformation with
$\gamma_5$. However, their Yukawa interactions with the scalar fields
inevitably become non-linear. Such a field redefinition does not make
sense on the lattice because of the singularity at the momentum region
where the species doublers appear.  

We have two ways in realization of chiral symmetry, one with non-linear
symmetry transformations, and another with linear transformations.  
Even for the latter, the Wilson actions which are solutions to the WT
identities are non-linear in scalar fields.  

This paper is organized as follows. We briefly summarize a functional
integral method for derivation of the WT identities for a global
symmetry in the next section, and obtain a generalized GW relation for a
$\sigma$ model in section 3. We give a family of non-perturbative
solutions, and discuss some concrete solutions in section 4. We then
discuss that the family of the Wilson actions constructed from the
solutions given in section 4 can be reduced to a family of the Wilson
actions for which chiral symmetry is linearly realized.
The last section is devoted to summary and
outlook.

\section{Derivation of the WT identities for a global symmetry}
      
Let us consider a generic renormalizable theory with 
action $S_{0}[\phi]$ in 4-dimensional Euclidean space. 
It is a functional of some fields that are collectively denoted
by $\phi^{A}$. The index $A$ represents the Lorentz indices
of vector fields, the spinor indices of the fermions, and/or indices
distinguishing different types of generic fields.  The Grassmann parity
for $\phi^{A}$ is expressed as $\ep(\phi^{A})=\ep_{A}$, so that
$\ep_{A} =0$ if the field $\phi^{A}$ is Grassmann even (bosonic) and
$\ep_{A} =1$ if it is Grassmann odd (fermionic).

In order to regularize the theory, we introduce an IR momentum cutoff
$\Lambda$ and  through a positive
function that behaves as
\begin{eqnarray}
 \quad K\Bigl(\frac{p^{2}}{\Lambda^{2}}\Bigr)\approx \quad  \left\{
		\begin{array}{ll}
		 1 & (p^2 < \Lambda^2)~, \\
		 0 & (p^2 >  \Lambda^2)~.
		\end{array}
               \right.
\label{cutoff-func}
\end{eqnarray}
We also introduce a UV cutoff $\Lambda_{0} > \Lambda$ and  
$K_{0}(p)\equiv K(p^{2}/\Lambda_{0}^{2})$.

Introducing sources $J_A$ for the fields $\phi^{A}$, the generating functional is written as
\begin{eqnarray}
{\cal Z}_{\phi}[J] = \int {\cal D} \phi 
\exp\left(-{\cal S}[\phi ; \Lambda_{0}]+ K_{0}^{-1} J \cdot \phi
       \right ),
\label{part-func1}
\end{eqnarray}
where the action ${\cal S}$ defined at the scale $\Lambda_{0}$ is
written as the sum of the kinetic and interaction terms
\begin{eqnarray}
{\cal S}[\phi; \Lambda_{0}] 
= \frac{1}{2}\phi \cdot K_{0}^{-1} D_{0} \cdot \phi + 
{\cal S}_{I}[\phi; \Lambda_{0}]\,.
\label{micro-action}
\end{eqnarray}
Here we use the matrix notation in momentum space:
\begin{eqnarray}
J \cdot \phi&=& \int \frac{d^{4}p}{(2\pi)^4}J_{A}(-p)\phi^{A}(p)\,, \nn\\
\phi \cdot  D  \cdot \phi &=& \int \frac{d^{4}p}{(2\pi)^4} 
\phi^{A}(-p) D_{AB}(p) \phi^{B}(p)\,.
\label{cond-not}
\end{eqnarray}

In the ERG approach , one introduces IR fields $\Phi^{A}$ whose action, 
the Wilson action $S[\Phi]$, is given as a functional integral of the original 
UV fields $\phi^{A}$. This can be done to rewrite 
 the partition function \bref{part-func1} up to field and source independent 
constant as
\begin{eqnarray}
{\cal Z}_{\phi}[J] &=& \int {\cal D} \phi ~ {\cal D} \Phi 
\exp -\frac{1}{2}\Bigl(\Phi -f~\phi - J \cdot (K\alpha)^{-1}\Bigr)
\cdot \alpha \cdot \Bigl(\Phi -f~\phi - (-)^{\ep_{J}}(K\alpha)^{-1}\cdot J  \Bigr) \nn\\
&& \times\exp\left(-{\cal S}[\phi ; \Lambda_{0}]+ K_{0}^{-1} J \cdot \phi
       \right )\,,
\end{eqnarray}
where $f=K/K_{0}$, and matrix $\ds{\alpha=\alpha_{AB}}$ is the kernel of the gaussian 
integral over the IR fields. We obtain the following expressions for the Wilson action:
\begin{eqnarray}
{\cal Z}_{\phi}[J] &=& N_{J} {Z}_{\Phi}[J] \,,\nn\\
{Z}_{\Phi}[J] &=& \int {\cal D} \Phi 
\exp\left(-{S}[\Phi; \Lambda]+ J \cdot K^{-1}\Phi
       \right )\,,
\label{part-func2}\\
\exp \left( -S[{\Phi}; \Lambda]\right) &=&  \int{\cal D}\phi
\exp -\left\{\frac{1}{2}(\Phi -f~\phi)
\cdot \alpha \cdot (\Phi -f~\phi) + {\cal S}[\phi ; \Lambda_{0}]\right\}\,,
\nn
\end{eqnarray} 
where the normalization factor $N_{J}$ is given by
\begin{eqnarray}
\ln N_{J} &=& -\frac{(-)^{\ep_{A}}}{2} 
J_{A} K^{-2} \left(\alpha^{-1}\right)^{AB} J_{B}~.
\label{normalization}
\end{eqnarray}


We discuss symmetry properties of the Wilson action. Consider a change
of variables defined at the UV scale $\Lambda_0$:
\begin{eqnarray}
\phi^{A} \rightarrow \phi^{\prime A} =\phi^{A} +  \delta \phi^{A}~,\qquad
\delta \phi^{A} = {\cal R}^{A}[\phi]\,.
\end{eqnarray}
The generating functional (\ref{part-func1}) is invariant under the
change of the integration variable. It gives the relation
\begin{eqnarray}
\int {\cal D}\phi 
\Bigl(K_0^{-1} J \cdot \delta \phi - \Sigma [\phi; \Lambda_0]\Bigr) 
~\exp\left(-{\cal S}[\phi ; \Lambda_{0}]
+K_{0}^{-1} J \cdot \phi\right) = 0  \,,
\label{equality}
\end{eqnarray}
where $\Sigma [\phi; \Lambda_0]$ is {\it the WT operator} given as
\begin{eqnarray}
\Sigma[\phi; \Lambda_{0}] 
 \equiv 
\frac{\pa^{r} {\cal S}}{\pa \phi^{A}} \delta \phi^{A} 
- (-)^{\ep_{A}}\frac{\pa }{\pa \phi^{A}} \delta \phi^{A}~. 
\label{WTop1}
\end{eqnarray} 
$\Sigma [\phi, \Lambda_0]$ is the sum of the change of the action ${\cal S}[\phi ; \Lambda_{0}]$
\begin{eqnarray}
\delta {\cal S} = \frac{\pa^{r} {\cal S}}{\pa \phi^{A}}
\delta \phi^{A}\,,
\label{delta-cal-S}
\end{eqnarray} 
and that of the functional measure ${\cal D} \phi$
\begin{eqnarray}
\delta \ln {\cal D} \phi = (-)^{\ep_{A}}
 \frac{\pa^{r} }{\pa \phi^{A}} \delta \phi^{A}\,.
\label{measure}
\end{eqnarray} 

The relation (\ref{equality}) may be rewritten as
\begin{eqnarray}
\left\langle\Sigma [\phi ; \Lambda_{0}]\right\rangle_{\phi, ~K_{0}^{-1}J}
&=&
K_0^{-1} J \cdot \left<\delta \phi \right\rangle_{\phi,
~K_{0}^{-1}J} \nonumber\\
&=&
K_0^{-1} J \cdot \left<{\cal R} [\phi ; \Lambda_{0}]\right\rangle_{\phi, ~K_{0}^{-1}J}
\nn\\
&=&
K_0^{-1} J \cdot {\cal R} [K_0 \partial_J^l ; \Lambda_{0}]~{\cal Z}_{\phi}[J]\nn\\
&=& N_{J} \biggl\{N_{J}^{-1} \Bigl(K_0^{-1} J \cdot {\cal R} [K_0 \partial_J^l ; \Lambda_{0}]~N_{J}
\Bigr) \nn\\
&& + K_0^{-1} J \cdot {\cal R} [K_0 \partial_J^l ; \Lambda_{0}]~\biggr\}~Z_{\Phi}[J]\nn\\
&=& \left\langle\Sigma [\Phi ; \Lambda]\right\rangle_{\Phi, ~K^{-1}J} \,.
\label{equality2}
\end{eqnarray}

In this paper we assume that the theory given by \bref{part-func1} and
\bref{micro-action} admits a linear global symmetry described by
\begin{eqnarray} 
\Sigma [\phi ; \Lambda_{0}] &=&  \delta {\cal S}
= \frac{\pa^{r} {\cal S}}{\pa \phi^{A}}~\delta \phi^{A} =0 \,,\nn\\
\delta \phi^{A} &=& {\cal R}^{A}_{~~B}[\Lambda_{0}] \phi^{B}\,,
\end{eqnarray}
where ${\cal R}^{A}_{~~B}$ do not depend on the fields.  Renormalization
affects and deforms this symmetry.  To see this, 
we compute
\begin{eqnarray}
    K_{0} \frac{\pa^{l}}{\pa J_{A}} {\cal Z}_{\phi}[J]
    &=& K_{0}\frac{\pa^{l}}{\pa J_{A}}
    N_{J}Z_{\Phi}[J]\nn\\
    &=& N_{J} \left[(-)^{\ep_{A} +1} K_{0}
        K^{-2} \left(\alpha^{-1}\right)^{AB}J_{B} +
        K_{0}\frac{\pa^{l}}{\pa J_{A}}\right]   Z_{\Phi}[J]
    \nn\\
    &=& N_{J} \left\langle f^{-1}\left\{\Phi^{A} - \left(\alpha^{-1}\right)^{AB}
        \frac{\pa^{l} S }{\pa
	\Phi^{B}}\right\}\right\rangle_{\Phi,K^{-1}J}\,,
\label{CO1}
\end{eqnarray}
which can be used to obtain the WT identity for the IR fields
\begin{eqnarray}
\Sigma[\Phi; \Lambda] 
&=& \frac{\pa^{r} {S}}{\pa \Phi^{A}} \delta \Phi^{A} 
- (-)^{\ep_{A}}\frac{\pa }{\pa \Phi^{A}} \delta \Phi^{A}~,\nn\\
\delta \Phi^{A} 
&=& {\cal R}^{A}_{~~B}[\Lambda_{0}] 
\left\{\Phi^{A} 
- \left(\alpha^{-1}\right)^{AB}
\frac{\pa^{l} S }{\pa \Phi^{B}}\right\}\,.
\label{WTop2}
\end{eqnarray}

We consider in the next section non-perturbative realization of chiral symmetry 
for a linear sigma model in the ERG approach.

\section{The WT identities for chiral symmetry}  

The model we deal with in this paper consists of a Dirac field,
$\psi,~\bar\psi$, and complex scalar fields, $\phi,~\phi^{\dagger}$,
collectively denoted as $\phi^{A}
=\{\psi,~\bar\psi,~\phi,~\phi^{\dagger}\}$.  The UV action is given by
\begin{eqnarray}
{\cal S}[\phi] &=& 
{\cal S}[\bar\psi, \psi, \theta] + {\cal S}[\phi, \phi^{\dagger}] \,,\nn\\
 {\cal S}[\bar\psi, \psi, \theta]  &=&
\int_{p,q}~ \bar\psi(-p) \biggl[ K_{0}(p)^{-1}\Slash{p}\delta(p-q) + 
\theta(p-q)\biggr]\psi(q) \,,\nn\\ 
{\cal S}[\phi, \phi^{\dagger}] &=& \int_{p} K_{0}(p)^{-1}
\phi^{\dagger}(-p) p^{2} \phi(p)
 + {\cal S}_{I}[\phi, \phi^{\dagger}]\, ,
\label{UVaction}
\end{eqnarray}
where
\begin{eqnarray}
\theta(p) &=& P_{+} \phi(p) + P_{-} \phi^{\dagger}(p) \,,\nn\\
P_{\pm} &=& \frac{1 \pm \gamma_{5}}{2}\, .
\end{eqnarray}
The action \bref{UVaction} is invariant under chiral transformation
\begin{eqnarray} 
\delta \psi(p) &=& i \varep  \gamma_{5} \psi(p)\,, \qquad
\delta \bar\psi(-p) = \bar\psi(-p)  i \varep  \gamma_{5} \,,\nn\\
\delta \phi (p) &=& -2i  \varep  \phi (p)\,, \qquad
\delta \phi^{\dagger} (-p) = 2i  \varep  \phi^{\dagger} (-p)\,,
\end{eqnarray}
with a Grassmann even parameter $\varep$. Since we wish to consider 
a non-trivial realization of chiral symmetry, we will take 
a chiral non-invariant blocking kernel for the Dirac field 
\begin{eqnarray} 
\alpha_{D} (p) =  \frac{M~K_{0}(p)}{K(p) (K_{0}(p)-K(p))}\,,
\label{defalpha}
\end{eqnarray}
where $M$ is a constant with mass dimension 1. For the scalar field, 
, we take a blocking kernel 
\begin{eqnarray} 
\alpha_{S} (p) =  \frac{K_{0}(p) p^{2}}{K(p) (K_{0}(p)-K(p))}\,.
\end{eqnarray}

For the chiral symmetry, the WT identity \bref{WTop2} 
for the IR fields
$\Phi^{A} =\{ \Psi,~\bar\Psi,~\varphi,~\varphi^{\dagger}\}$
becomes
\begin{eqnarray}
\Sigma[\Phi] &=& \int_{p} \biggl[
\frac{\pa^{r} S}{\pa \Psi(p)} i \varep 
\gamma_{5} \biggl\{\Psi(p)- \alpha_{D}^{-1}(p) \frac{\pa^{l} S}{\pa \bar\Psi(-p)}\biggr\}
+ \biggl\{\bar\Psi(-p) - \alpha_{D}^{-1}(p)\frac{\pa^{r} S}{\pa \Psi(p)}
\biggr\}
i \varep \gamma_{5} \frac{\pa^{l} S}{\pa \bar\Psi(-p)} \nn\\
&& - \frac{\pa^{r} }{\pa \Psi(p)}2i\varep  \gamma_{5} \alpha_{D}^{-1}(p) \frac{\pa^{l} S}{\pa \bar\Psi(-p)} + \frac{\pa S}{\pa \varphi(p)}(-2i)
\biggl\{\varphi(p)- \alpha^{-1}_{S}(p) \frac{\pa S}{\pa \varphi^{\dagger}(-p)}\biggr\}\nn\\
&&+ \biggl\{\varphi^{\dagger}(-p) - \alpha^{-1}_{S}(p)\frac{\pa S}{\pa \varphi(p)} \biggr\}(2i)  \frac{\pa S}{\pa \varphi^{\dagger}(-p)}
\biggr]\nn\\
&=& \int_{p} \biggl[
\frac{\pa^{r} S}{\pa \Psi(p)} i \varep \gamma_{5} \biggl\{\Psi(p)-2 
\alpha_{D}^{-1}(p) \frac{\pa^{l} S}{\pa \bar\Psi(-p)}\biggr\}
+ \bar\Psi(-p) i \varep \gamma_{5} \frac{\pa^{l} S}{\pa \bar\Psi(-p)}\nn\\
&&\quad - \frac{\pa^{r} }{\pa \Psi(p)}2i\varep  \gamma_{5} \alpha_{D}^{-1}(p) \frac{\pa^{l} S}{\pa \bar\Psi(-p)}
 + \frac{\pa S}{\pa \varphi(p)}(-2i) \varphi(p) 
+ \varphi^{\dagger}(-p)(2i) \frac{\pa S}{\pa \varphi^{\dagger}(-p)}\biggr]
\nn\\
&=& \quad 0\,.
\label{WT-chiral1}
\end{eqnarray}
Note that the deformation of chiral symmetry only appears in the
fermionic sector. Furthermore, 
We define the chiral transformation in
such a way that non-trivial deformation only appears in $\delta \Psi$: 
\begin{eqnarray}
\delta \Psi(p) &=& i \varep \gamma_{5} \biggl\{\Psi(p)
-2 \alpha_{D}^{-1}(p) \frac{\pa^{l} S}{\pa \bar\Psi(-p)}\biggr\}\,,\qquad
\delta \bar\Psi(-p) = \bar\Psi(-p) i \varep \gamma_{5} \,,\nn\\
\delta \varphi (p) &=& -2i  \varep  \varphi (p)\,, \qquad
\delta \varphi^{\dagger} (-p) = 2i  \varep  \varphi^{\dagger} (-p)\,.
\label{chiral-tr3}
\end{eqnarray}

In the following section, we try to construct solutions to
\bref{WT-chiral1} regarding it as purely algebraic equation.
Therefore,
we are not confined ourselves to the specific solution of 
Wilson action obtained explicit
integration of the blocked UV action.   
For simplicity, we shall use $\alpha=\alpha_{D}$ below.

\section{Solutions to the WT identities}
\subsection{GW relation in free-field theory}
Let us first consider a free-field Wilson action $S_{0}$ for the IR fields 
$\Phi^{A} =\{ \Psi,~\bar\Psi,~\varphi,~\varphi^{\dagger}\}$
\begin{eqnarray} 
 S_{0}[\Phi]&=& 
S_{0}[\bar\Psi, \Psi] + S_{0}[\varphi, \varphi^{\dagger}] \,,\nn\\
S_{0}[\bar\Psi, \Psi] &=& \int_{p}  \bar\Psi(-p) {\cal D}_{0}(p)
\Psi(p) \,,\nn\\
 S_{0}[\varphi, \varphi^{\dagger}]&=& \int_p
K^{-1}(p) \varphi^{\dagger}(-p)  ~p^{2}
\varphi(p)\, .
\label{Wilson-kinetic}
\end{eqnarray}
Using \bref{chiral-tr3}, we obtain chiral transformation for free-field theory, 
\begin{eqnarray} 
\delta \Psi(p) &=& i \varep \gamma_{5} \left(1 -2\alpha^{-1}{\cal D}_{0}\right)(p)\Psi(p)\,, \qquad
\delta \bar\Psi(-p) = \bar\Psi(-p) i \varep \gamma_{5} \,,\nn\\
\delta \varphi(p) &=& -2i \varep \varphi(p)\,, \qquad 
\delta \varphi^{\dagger}(-p) = 2i \varep \varphi^{\dagger}(-p) \,.
\end{eqnarray}
The WT identity for the free-field action
\begin{eqnarray}
\Sigma[\Phi] =\frac{\pa^{r} S_{0}}{\pa \Phi^{A}} \delta \Phi^{A} =0
\end{eqnarray}
leads to the GW relation
\begin{eqnarray}
\{ \gamma_{5},~{\cal D}_{0}\} = 2  {\cal D}_{0} \gamma_{5}\alpha^{-1} 
{\cal D}_{0}\,.
\label{gw-free}
\end{eqnarray}
For our purpose, we don't need to specify the Dirac operator ${\cal D}_0$
for free Dirac fields, and we only assume that it satisfies the GW
relation \bref{gw-free}. 

\subsection{The generalized GW relation for interacting theory}
We next include interaction terms of the Dirac fields with the 
scalars expressed by 
\begin{eqnarray}
\vartheta (p) \equiv  P_{+} \varphi(p) +
		     P_{-}\varphi^{\dagger}(p) \,,
\label{theta}
\end{eqnarray}
where $P_{\pm} = (1 \pm \gamma_{5})/2$. Our basic assumption is 
that IR
action is bilinear in the fermionic fields:
\begin{eqnarray}
S[\Phi] &=& S_{1}[\bar\Psi, \Psi, {\vartheta}] + 
S_{2}[\varphi, \varphi^{\dagger}] \,,\nn\\
S_{1}[\bar\Psi, \Psi, \cal{\vartheta}] &=& 
\int_{p,q} \bar\Psi(-p) {\cal D}(p, q)
\Psi(q) \,,\nn\\
 S_{2}[\varphi,
 \varphi^{\dagger}] &=& S_{0}[\varphi, \varphi^{\dagger}] 
 + S_{I}[\varphi, \varphi^{\dagger}] + S_{\rm counter}[{\cal \vartheta}]\, ,
\label{IRaction}
\end{eqnarray}
where $S_{0}[\varphi, \varphi^{\dagger}]$ is the free-field action 
\bref{Wilson-kinetic} for scalars. $S_{0} + S_{I}$ is chiral invariant,
while $S_{\rm counter}[{\vartheta}]$ is not. The latter is determined below.
The Dirac operator ${\cal D}$ is assumed to take the form 
\begin{eqnarray}
{\cal D}(p, q) = {\cal D}_{0}(p) \delta(p-q) + 
\eta(p)
{\cal V}(p,q)
\eta(q)
\,,
\label{Dirac-op}
\end{eqnarray}
where ${\cal D}_{0}$ is the Dirac operator which satisfies the GW
relation \bref{gw-free} for free-fields, and 
\begin{eqnarray}
\eta(p) = 1-\alpha^{-1}(p){\cal D}_0(p) \, .
\label{eta}
\end{eqnarray}
 ${\cal V}$ is some functional of ${\vartheta}$. We try to find
 most general solutions 
to \bref{WT-chiral1}, assuming the IR action is bilinear in the Dirac fields.

For the action \bref{IRaction}, we divide the WT identity
\bref{WT-chiral1} into two parts:
\begin{eqnarray}
\Sigma[\Phi] &=& \Sigma_{1}[\Phi] + \Sigma_{2}[\Phi] \,,
\nn\\
\Sigma_{1}[\Phi] 
&=& \int_{p} \biggl[
\frac{\pa^{r} S_{1}}{\pa \Psi(p)} i \varep \gamma_{5} \biggl\{\Psi(p)-2 
\alpha^{-1}(p) \frac{\pa^{l} S_{1}}{\pa \bar\Psi(-p)}\biggr\}
+ \bar\Psi(-p) i \varep \gamma_{5} \frac{\pa^{l} S_{1}}{\pa \bar\Psi(-p)}\nn\\
&& + \frac{\pa S_{1}}{\pa \varphi(p)}(-2i) \varphi(p)
+\varphi^{\dagger}(-p)(2i) \frac{\pa S_{1}}
{\pa \varphi^{\dagger}(-p)}
 \biggr] \,,
\nn\\
\Sigma_{2}[\Phi] &=& \int_{p}
\biggl[- \frac{\pa^{r} }{\pa \Psi(p)}2i\varep  \gamma_{5} 
\alpha^{-1}(p) \frac{\pa^{l} S_{1}}{\pa \bar\Psi(-p)}
 + \frac{\pa S_{2}}{\pa \varphi(p)}(-2i) \varphi(p) \nn\\
&&
+ \varphi^{\dagger}(-p)(2i) \frac{\pa S_{2}}{\pa
\varphi^{\dagger}(-p)}\biggr]\, .
\label{WT-chiral2}
\end{eqnarray}
Since $\Sigma_{1}$ is bilinear in the Dirac fields while $\Sigma_{2}$
only contains scalar fields, each should separately vanish
\begin{eqnarray}
\Sigma_{1}[\Phi] &=& 0 \,,\nn\\
\Sigma_{2}[\Phi] &=& 0\, .
\label{WT-chiral3}
\end{eqnarray}

Let us first consider the WT identity $\Sigma_{1}[\Phi] = 0$.
It leads to a generalized GW relation 
\begin{align}
\big\{\gamma_5,\mathcal{D}(p,q)\big\} 
= 
2\int_k\mathcal{D}(p,k)\gamma_5\alpha^{-1}(k)\mathcal{D}(k,q)-(i\epsilon)^{-1}\delta\mathcal{D}(p,q)\,.
\label{gwRel}
\end{align} 
This generates an equation for chiral transformation of ${\cal V}$
\begin{align}
\delta{\cal V}(p,q) 
&= 
-i\epsilon\bigg[
\big\{\gamma_5,{\cal V}(p,q)\big\} 
- 2\int_k{\cal V}(p,k)\eta(k) \gamma_5\alpha^{-1}(k)
\eta(k)
{\cal V}(k,q) \bigg] \notag \\
&= 
-i\epsilon\bigg[
\big\{\gamma_5,{\cal V}(p,q)\big\} 
- \int_k{\cal V}(p,k) \alpha^{-1}(k)
\big\{
\gamma_5,~  \eta(k)
\big\}
{\cal V}(k,q) \bigg] \,,
\label{defdeltaTheta}
\end{align}
where we have used the relation
\begin{align}
2~\eta(p) \gamma_5 \eta(p) = \big\{
\gamma_5,~  \eta(p)
\big\}\, ,
\label{eta-rel}
\end{align}
which follows from the GW relation \bref{gw-free} for free-fields.

In order to solve \bref{defdeltaTheta}, we introduce a composite object
$\Theta(p)$ which is assumed to obey the same chiral transformation 
as $\vartheta$:
\begin{align}
\delta{\Theta(p,q)}=-i\epsilon \{\gamma_5,~ {\Theta(p,q)} \}
\label{defTheta0}
\end{align}
Using this object, we express ${\cal V}$ as
\begin{align}
{\cal V}(p,q) & = \bigg[{\Theta}(p,q) - \int_{k} {\Theta}(p,k)
 B(k){\Theta}(k,q) \nn\\
& + \int_{k,l} {\Theta}(p,k) B(k) {\Theta}(k,l) B(l) {\Theta}(l,q) 
+ \cdots\bigg]  \nn\\
& \equiv  \Big({\Theta}\frac{1}{1+B{\Theta}}\Big)(p,q) 
= \Big(\frac{1}{1+{\Theta} B} {\Theta}\Big)(p,q)  \, .
\label{defTheta}
\end{align}
The form of ${\cal V}$ is suggested in  ref.~\citen{Igarashi-So-Ukita}
\footnote{We may define ${\cal V}(p,q) = A(p) 
\Big({\Theta}\frac{1}{1+B{\Theta}}\Big)(p,q) C(q)$ by introducing
$A$ and $C$ which commute with $\gamma_5$. 
However, $A, C$ can be absorbed by redefinition of $\Theta$ and $B$ as 
$\Theta(p,q) \to \Theta'(p,q) =A(p) \Theta(p,q) C(q)$ and $B(p) \to B'(p) =
C^{-1}(p) B(p) A^{-1}(p)$.}.  
We discuss in the next section how this form emerges from
the point of view of linear realization of chiral symmetry.

Using \bref{defTheta0}, and
taking the average of two expressions for $\delta {\cal V}$ obtained from 
\bref{defTheta}, we find 
\begin{align}
\delta{\cal V}(p,q) 
&= 
-i\epsilon\bigg[
\gamma_5  {\cal V}(p,q) + {\cal V}(p,q) \gamma_5  \nn\\
& - \int_k{\cal V}(p,k) \big\{\gamma_5\,,\,B(k)\big\}{\cal V}(k,q)\bigg] \, .
\label{deltaTheta2}
\end{align}
Comparing \bref{defdeltaTheta} with \bref{deltaTheta2}, we have the conditions
\begin{gather}
\big\{ \gamma_5 \,,\, B \big\}  = 2 
\alpha^{-1} \eta \gamma_5 \eta\,.
\label{eqA}
\end{gather}
Note that the composite object $\Theta$ is subjected to chiral
transformation \bref{defTheta0}, otherwise arbitrary. It can be expanded
in terms of the original field $\vartheta$ as
\begin{align}
\Theta(p, q) = \beta_{1}(p) \vartheta(p-q) \xi_{1}(q) + 
\int_{k,l} \beta_{2}(p) \vartheta(p-k) \kappa(k,l)_{2} \vartheta(l-q) 
 \xi_{2}(q) + \cdots\,,
\label{expTheta}
\end{align}
where the coefficients functions $\beta$'s, $\xi$'s and the matrices
$\kappa$'s are commuting or anticommuting functions of momentum
variables. They are subjected to the boundary condition $\Theta
\rightarrow \vartheta$ as $\Lambda \rightarrow \Lambda_{0}$, otherwise
arbitrary.  Thus, we conclude that \bref{defTheta} 
is the most general set of  
solutions to \bref{defdeltaTheta} for $B$ satisfying \bref{eqA}.
The Dirac
operator with this ${\cal V}$ is given by
\begin{align}
{\cal D}(p, q) &= {\cal D}_{0}(p) \delta(p-q) + 
\eta(p)
\Big({\Theta}\frac{1}{1+B{\Theta}}\Big)(p,q)
\eta(q)
\, .
\label{Dirac-op1}
\end{align}
We may rewrite it as
\begin{align}
\mathcal{D}(p,q) 
&= 
\mathcal{D}_0(p)\delta(p-q) 
+ \gamma_5 \eta^{-1}(p)
\gamma_5
{\cal V}'(p,q) \eta(q) \,,
\label{opDirac}
\end{align}
where
\begin{align}
{\cal V}'(p,q) 
= 
\gamma_5 \eta(p) \gamma_5 \eta(p)
\Big(\Theta\frac{1}{1+B\Theta}\Big)(p,q) \, .
\label{Theta'1}
\end{align}
The chiral transformation of ${\cal V}'$ is given by
\begin{align}
\delta{\cal V}'(p,q) 
= 
-i\epsilon\bigg(
\big\{\gamma_5\,,\,{\cal V}'(p,q)\big\}  
- 2\int_k{\cal V}'(p,k)\gamma_5\alpha^{-1}(k){\cal V}'(k,q)
\bigg) \,.\label{deltaTheta1}
\end{align}
\\
\\
Let us discuss the WT identity $\Sigma_{2}[\Phi]=0$. 
The WT operator $\Sigma_{2}[\Phi]$ contains the Jacobian factor
$J $  associated with the chiral transformation for the Dirac fields $\delta
\Psi$. It is given by
\begin{align}
J &= - \int_{p}
\frac{\pa^{r} }{\pa \Psi(p)}2i\varep  \gamma_{5} 
\alpha^{-1}(p) \frac{\pa^{l} S_{1}}{\pa \bar\Psi(-p)}
 = -2i \varep  {\rm Tr}\Big[\gamma_{5} \alpha^{-1}
{\cal D}\Big] \nn\\
&=  -2i \varep  {\rm Tr}\Big[\gamma_{5}  \alpha^{-1}
{\cal V}'\Big]\, .
\label{Jacobian}
\end{align}
where ${\rm Tr}$ should be taken over the spinor and the momentum.

In order to cancel this contribution, the scalar sector action $S_{2}$
should contain a counter term
\begin{align}
 S_{\rm counter}[{{\cal V}'}] =  {\rm Tr} \log\Big[1 - \alpha^{-1}
{\cal V}'\Big] 
\,.
\label{counter0}
\end{align}
It reduces to 
\begin{align}
 S_{\rm counter}[{{\cal V}'}] &=  {\rm Tr} \log\Big[1 - \alpha^{-1}
{\cal V}'\Big] \nn\\
&= 
\text{Tr}\log\bigg(1-\frac{\gamma_5}{2}\big\{\gamma_5\,,\,B\big\}
\Theta\frac{1}{1+B\Theta}\bigg) \notag\\
&= 
\text{Tr}\log\bigg(1-B \Theta\frac{1}{1+B\Theta} 
+ \frac{\gamma_5}{2}\big[\gamma_5\,,\,B\big]
\Theta\frac{1}{1+B\Theta}\bigg) \notag\\
&= 
\text{Tr}\log\bigg[\Big(1 +
 \frac{\gamma_5}{2}[\gamma_5,B]\Theta\Big)\frac{1}{1+B\Theta}\bigg]
\, .
\label{couter-rel}
\end{align}
We define 
\begin{align}
S_{\rm counter}[{\Theta}] = - 
{\rm Tr} \log (1 + B {\Theta})\, .
\end{align}
Note that the difference 
$S_{\rm counter}[{{\cal V}'}] -S_{\rm counter}[{\Theta}] $ is chiral
invariant:
\begin{align} 
\delta \Big(S_{\rm counter}[{{\cal V}'}]  -  S_{\rm counter}[{\Theta}]\Big) = 
\delta\text{Tr}\log\Big(1+\frac{\gamma_5}{2}\big[\gamma_5\,,\,B\big]\Theta\Big) 
= 0\, .
\end{align}
Since the counter action is determined up to a chiral invariant term, we
may take $S_{\rm counter}[{\Theta}]$ as a counter action:
\begin{align}
\Sigma_{2}[\Phi] = J + \delta S_{\rm counter}[{{\cal V}'}]=
J + \delta S_{\rm counter}[{\Theta}] = 0\, .
\end{align}
In this way we have obtained generic IR action which solves the WT identity
$\Sigma[\Phi]=0$.
\\
\\
We shall discuss a set of solutions obtained 
by taking appropriate $\Theta$, and special functions for $B$.

\vspace{0.5cm} 

\noindent
1) $\Theta(p,q) =\eta^{-1}(p) \gamma_5 \eta^{-1}(p) \gamma_5 \vartheta(p-q) 
, ~B =\alpha^{-1}\gamma_5 \eta \gamma_5 \eta$\,.

It gives the Dirac operator
\begin{align}
\mathcal{D}(p,q) 
= 
\mathcal{D}_0(p)\delta(p-q) 
+ \gamma_5 \eta^{-1}(p)
\gamma_5
\Big(\vartheta\frac{1}{1+\alpha^{-1}\vartheta}\Big)(p,q)
\eta(q)
\, .
\label{Dirac1}
\end{align}

\noindent
2) $\Theta(p,q) =\vartheta(p-q)  \gamma_5 \eta^{-1}(q)  \gamma_5 \eta^{-1}(q)
, ~B =\alpha^{-1}  \eta \gamma_5 \eta\gamma_{5}$

It gives the Dirac operator where the factors $\eta$ and $\gamma_5 \eta^{-1}
\gamma_5$ appeared in \bref{Dirac1} is exchanged:
\begin{align}
\mathcal{D}(p,q) 
= 
\mathcal{D}_0(p)\delta(p-q) 
+ \eta(p)
\Big(\vartheta\frac{1}{1+\alpha^{-1}\vartheta}\Big)(p,q)
\gamma_5\eta^{-1}(q) \gamma_5
\, .
\label{Dirac2}
\end{align}

The set of the solutions 1) and 2) is   
the continuum analog of the one given in ref.~\citen{Igarashi-So-Ukita} on the
lattice.  We stress that the factor $\eta$ is
equal to $1$ at $\Lambda= \Lambda_{0}$ and it is regular for $\Lambda <
\Lambda_{0}$  in continuum theory considered here. On the lattice, the
corresponding factor vanishes at the momentum region where species
doublers appear. For both of 1) and 2), the counter action
\bref{counter0} reduces to
\begin{align}
 S_{\rm counter}[{\vartheta}] = - 
{\rm Tr} \log (1 + \alpha^{-1} {\vartheta})\,.
\label{counter1}
\end{align}

\noindent
3) $\Theta(p,q) = f^{-1}(p) \vartheta(p-q) f^{-1}(q), ~
B= \alpha^{-1} \eta $ \,, \\
This solution which gives the Dirac operator
\begin{align}
\mathcal{D}(p,q)
= 
\mathcal{D}_0(p)\delta(p-q) + 
f^{-1}(p)\eta(p) \bigg( \vartheta
\frac{1}{1+f^{-2}\alpha^{-1}\eta\vartheta}\bigg)(p,q) f^{-1}(q)\eta(q) 
\label{irDiracOp} \, ,
\end{align}
is related to the Wilson action derived 
from explicit integration over the UV
Dirac fields. Actually, for the UV Dirac action 
\begin{align}
{\cal S}[\bar{\psi},\psi, \theta]
&= \int_{p,q}\bar{\psi}(-p)D(p,q)\psi(q) \,,\nn\\
D(p,q) &= 
K_0^{-1}(p)\cancel{p}\,\delta(p-q)+ 
\theta(p-q)
\, ,
\end{align}
we may integrate the blocked action:
\begin{align}
e^{-S[\bar{\Psi},\Psi, \theta] } 
&= 
\int\mathcal{D}\bar{\psi}\mathcal{D}\psi \exp\bigg[
-\int_{p,q}\Big(\bar{\Psi}(-p)-f(p)\bar{\psi}(-p)\Big)\alpha(p)\delta(p-q)
\Big(\Psi(q)-f(q)\psi(q)\Big) \notag\\
&\hspace{2.7truecm} 
-\int_{p,q}\bar{\psi}(-p)D(p,q)\psi(q)
\bigg] \, .
\end{align}
The resulting Wilson action is given by
\begin{align}
S[\bar{\Psi},\Psi, \theta] 
&= 
\int_{p,q}\bar{\Psi}(-p)\mathcal{D}(p,q)\Psi(q) 
- \log \text{Det}(D+f^2\alpha) \,,\nn\\
\mathcal{D}(p,q) 
&= 
\alpha(p)\delta(p-q)-f(p)\alpha(p)(D+f^2\alpha)^{-1}(p,q)f(q)\alpha(q) \, .
\end{align} 
The $\log \text{Det}$ term is generated as the Lee-Yang term of the
functional integration over the UV Dirac fields.

We rewrite 
\begin{align}
(D+f^2\alpha) (p,q) &=\left( K_{0}^{-1}(p) \Slash{p} + f^{2}(p) 
\alpha(p)\right)\delta(p-q) + \theta(p-q) \nn\\
& \equiv \Delta^{-1}(p) \delta(p-q) + \theta(p-q)\, ,
\end{align}
and take its inverse as
\begin{align}
(D+f^2\alpha)^{-1} (p,q) &= 
\bigg[ \delta(p-q) - \Delta(p)\theta(p-q) + 
\int_{k}\Delta(p)\theta(p-k)\Delta(k)\theta(k-q) + \cdots
\bigg] \Delta(q) \\
&\equiv \Delta(p)\delta(p-q) - \Delta(p)
\Big( \theta \frac{1}{1+\Delta\theta} \Big)(p,q) \Delta(q) \, ,
\end{align}
We then have
\begin{align}
\mathcal{D}(p,q)
= 
\mathcal{D}_0(p)\delta(p-q) + 
f(p)\alpha(p)\Delta(p) \Big(\theta 
\frac{1}{1+\Delta\theta}\Big)(p,q) \Delta(q) f(q)\alpha(q) 
\label{irDiracOp1} \, ,
\end{align}
where the Dirac operator for free-fields takes the form 
\begin{align}
\mathcal{D}_0(p) &= 
 \alpha(p) - f(p)\alpha(p)\Delta(p)f(p)\alpha(p) \nn\\
&= f^{-1}(p)
\left\{\frac{M\Slash{p}}{(K_{0}(p)-K(p))\Slash{p} +  K(p) M }\right\}
\label{D0rel}
\end{align}
for $\alpha$ of \bref{defalpha}. Using
 \begin{align}
\eta = f^{2} \alpha \Delta\, ,
\end{align}
we find that the Dirac operator in \bref{irDiracOp} is obtained from that
of \bref{irDiracOp1} with a replacement of $\theta$ by $\vartheta$.    
Furthermore, the $\log \text{Det}$ (Lee-Yang) 
term in the action \bref{irDiracOp} is nothing but the counter action 
$S_{\rm counter}[{\vartheta}]$ when we identify $\theta=\vartheta$. 
The Wilson action with \bref{irDiracOp1} and $S_{\rm
counter}[{\vartheta}]$ is a perfect action if no blocking is performed
for the scalar fields.
For bilinear UV action for the Dirac fields, integration over
the UV Dirac fields automatically yields a solution to the WT identity
$\Sigma[\Phi] =0$ if we put  $\theta=\vartheta$ for scalar sector.

\section{Reduction to linear realization of chiral symmetry} 
\subsection{Chiral transformation with $\hat\gamma_5$}
In the previous section, we have obtained a certain family of the Wilson
actions which solves the WT identities. We have given some
concrete solutions choosing specific functions for $B$ and $\Theta$. 
Chiral transformations for each Wilson action are non-linear. 
In this section, we discuss that the family of the
Dirac actions $\bar\Psi {\cal D} \Psi$ with the Dirac operators 
\bref{Dirac-op1} reduces, by a field redefinition, to a family of the
Wilson actions for which chiral symmetry is linearly
realized. Note that our basic assumption that the Wilson
action is bilinear in the Dirac fields makes the linearization of
symmetry possible.

We begin with the expression for  the Dirac operator \bref{opDirac}, and 
introduce new Dirac field $\Psi'$ as
\begin{align}
\Psi'(p) = \int_{q} L^{-1}(p,q) \Psi(q)\, .
\label{defPsi'}
\end{align}
We require that its chiral transformation becomes
\begin{align}
\delta \Psi'(p) = i \varepsilon \hat\gamma_5 (p) \Psi'(p)
\label{defdelPsi'}
\end{align}
with
\begin{align}
\hat\gamma_5 (p)= \gamma_5 \left(1 - 2\alpha^{-1}(p) \mathcal{D}_0(p) \right) 
\, .
\end{align}
This $\hat\gamma_5$ is continuum counter part of the one extensively
used on the lattice. It 
satisfies $(\hat\gamma_5)^2
=1$, and can be used to define the chiral projection operators:
\begin{align}
{\hat P}_{\pm}(p) = \frac{1 \pm {\hat\gamma}_{5}(p)} {2}\,. 
\label{chiralpro-hat}
\end{align}

Our task is to find the operator $L$ in \bref{defPsi'}.  It follows from
\bref{defPsi'} and \bref{defdelPsi'} that it should obey the chiral
transformation
\begin{align}
\delta L(p,q) 
&=  
i\epsilon
\bigg(
\hat{\gamma}_5(p)L(p,q)-L(p,q)\hat{\gamma}_5(q) 
- 2\int_k\alpha^{-1}(p)\eta^{-1}(p)
\gamma_5
{\cal V}'(p,k) \eta(k)
L(k,q) \bigg)  \, .
\label{deltaL}
\end{align} 
As shown in Appendix A, a solution to this equation is given by
\begin{align}
L(p,q) 
= 
\eta^{-1}(p)
\frac{1}{1-\alpha^{-1}{\cal V}'}(p,q)
\eta(q)
\,.
\label{defL}
\end{align}
 
We next consider new Dirac operator for $\Psi'$:
\begin{align}
\mathcal{D}'(p,q) = \int_k  \mathcal{D}(p,k) L(k,q) \, .
\end{align}
It is straightforward to compute the r.h.s as shown in Appendix B:
\begin{align}
{\cal D}'(p,q)  
= {\cal D}_{0}(p)\delta(p-q)  + {\cal V}''(p,q)\, , 
\label{new-Dirac1}
\end{align} 
where 
\begin{align}
 {\cal V}''(p,q) &= 
\Big( {\cal V}' \frac{1}{1- \alpha^{-1} {\cal V}'}\Big)(p,q)\eta(q) \nn\\
&= 
\frac{1}{2} \alpha(p) \gamma_{5} \{\gamma_{5}, B(p)\} 
\biggl(\Theta \frac{1}{1+\frac{1}{2}\gamma_{5}[\gamma_{5},
 B]\Theta}\biggr)(p,q) \eta(q)
\, .
\label{Theta''}
\end{align} 
The chiral transformation of ${\cal V}''$ becomes linear
\begin{align}
\delta  {\cal V}''(p,q) =  -i \varepsilon \Big(\gamma_{5} {\cal V}''(p,q)
+  {\cal V}''(p,q) \hat\gamma_{5}(q) \Big) \,.
\end{align} 

The field redefinition \bref{defPsi'} with \bref{defL} induces a
Jacobian factor
\begin{align}
{\cal D}\Psi &= {\cal D}\Psi' \exp{\cal J} \,,\nn\\
{\cal J} &= {\rm Tr} \log \Big[ 1- \alpha^{-1} {\cal V}'\Big] \, ,
\end{align}
which exactly cancels the counter action \bref{counter0}.

In conclusion, the IR theory can be described by 
\begin{eqnarray}
{\cal D}\Psi' {\cal D}\Psibar {\cal D}\varphi {\cal D}\varphi^{\dagger}
\exp -S'[\Psi' ,\Psibar, \varphi, \varphi^{\dagger}] \,,
\end{eqnarray}
where the Wilson action is given by
\begin{eqnarray}
S'[\Psi' ,\Psibar, \varphi, \varphi^{\dagger}]&=&
\int_{p,q} \Psibar(-p){\cal D}'(p,q)\Psi'(q) + \int_{p} \varphi^{\dagger}(-p)
 K^{-1}(p) p^2 \varphi(p) + S_{I}[\varphi, \varphi^{\dagger}]  
\end{eqnarray}
with the Dirac operator given in \bref{new-Dirac1}.  
The action $S'$ is invariant under 
the chiral transformation
\begin{eqnarray}
\delta \Psi'(p) &=& i \varep \hat\gamma_{5}(p) \Psi'(p) \,,\nn\\
\delta \Psibar(-p) &=& \Psibar(-p) i \varep \gamma_{5} \,,\nn\\
\delta \varphi(p) &=& -2i \varep \varphi(p) \,,\nn\\
\delta \varphi^{\dagger}(-p)&=& 2i \varep \varphi^{\dagger}(-p)\,.
\end{eqnarray}
Chiral symmetry in this theory is simply expressed by the invariance of
the Wilson action $\delta S'=0$. 

It should be noticed that the composite object ${\cal V}''$ does not in
general commutes with $\gamma_{5}$, and 
cannot be
expressed by using the chiral projection operators. For the special
case, $B =\alpha^{-1}\gamma_5 \eta\gamma_5\eta$ and 
$\Theta(p,q) =\eta^{-1}(p) \gamma_5 \eta^{-1}(p)\gamma_5 \vartheta(p-q)$
, ${\cal V}''$ reduces to  
\begin{align}
{\cal V}''(p,q) = \vartheta(p-q) \eta(q) = 
P_{+} \varphi(p-q) 
{\hat P}_{+}(q) +P_{-} \varphi^{\dagger}(p-q) {\hat
 P}_{-}(q)\, ,
\end{align}
which is linear in $\varphi$ and $\varphi^{\dagger}$. This is the
continuum analog of the Yukawa couplings constructed on the lattice ref.~\citen{Igarashi-So-Ukita}.

\subsection{Chiral transformation with $\gamma_5$}

In the previous subsection, we have discussed a linear realization of
chiral symmetry using $\hat\gamma_5$.  Based on the ERG point of view, 
we discuss here that it is after
all possible to realize chiral symmetry as the standard way with the
constant $\gamma_5$ for the fermionic sector. Non-trivial modification
only appears in the Yukawa coupling with scalar fields. 

We begin with an IR action 
\begin{align}
S_{IR} = \int_{p,q}\bar\Psi(-p)\mathcal{D}(p,q)\Psi(q) +
 S_\text{counter}\,,
\label{IR-action1}
\end{align}
where the counter action is given by
\begin{align}
S_\text{counter} = \text{Tr}\log(1-\alpha^{-1}\mathcal{D})\,.
\label{counter-term1}
\end{align}

Consider the non-linear chiral transformation discussed in 4.2:
\begin{align}
\delta\Psi(p) &= i\epsilon\gamma_5\int_k\big(\delta(p-k)-2\alpha^{-1}(p)
\mathcal{D}(p,k)\big)\Psi(k)\nn\\
\delta \bar\Psi(-p) &= i\epsilon \bar\Psi(-p) \gamma_5 \,.
\label{non-linear-tr}
\end{align}
The Dirac operator $\mathcal{D}$ is subjected to the generalized GW
relation \bref{gwRel}
\begin{align}
\delta\mathcal{D}(p,q) = -i\epsilon\Big(\big\{\gamma_5,\mathcal{D}(p,q)\big\}
-2\int_k\mathcal{D}(p,k)\gamma_5\alpha^{-1}(k)\mathcal{D}(k,q)\Big)\,.
\label{gwRel1}
\end{align}
We perform a field redefinition
\begin{align}
\Psi'(p) = \int_k\big(\delta(p-k)-\alpha^{-1}(p)\mathcal{D}(p,k)\big)
\Psi(k) \,.
\label{field-red.2}
\end{align}
 In terms of new Dirac fields,  chiral transformation becomes the
 standard form 
 \begin{align}
\delta\Psi'(p) = i\epsilon\gamma_5\Psi'(p) \,,
\end{align}
and the new Dirac operator is given by
\begin{align}
\mathcal{D}^{\prime}(p,q) = \int_k \mathcal{D}(p,k)\frac{1}{1-\alpha^{-1}\mathcal{D}}(k,q)\,.
\label{D-relation}
\end{align}
Note that the Jacobian factor associated with the non-linear
transformation \bref{field-red.2} exactly cancels the counter action
$S_\text{counter}$. The GW relation \bref{gwRel1} is simplified as
\begin{align}
\big\{\gamma_5,\mathcal{D}'(p,q)\big\} = -(i\epsilon)^{-1}\delta
\mathcal{D}'(p,q)\,.
\label{gw-red1}
\end{align}
This is exactly the standard relation for chiral symmetry. If we take
the standard free Dirac operator such as $D_{0}(p) \sim \Slash{p}
$ and decompose the total Dirac operator $\mathcal{D}'$ as
\begin{align}
\mathcal{D}^\prime(p,q) = {D}_0 (p)\delta(p-q) +
 \Theta(p,q) 
\label{Dirac4}
\end{align}
where $\Theta(p,q) $ obeys
\begin{align}
\delta \Theta(p,q) = -i\epsilon\big\{\gamma_5, \Theta(p,q)\big\} \,.
\end{align}
Therefore, $\Theta$ can be expanded as \bref{expTheta}, and identifies
with that appeared in the previous section. It is obvious that the Dirac
operator \bref{Dirac4} provides the most general form of the solutions
to \bref{gw-red1}.

On the other hand, we may
also decompose the Dirac operator in \bref{IR-action1} as 
\begin{align}
\mathcal{D}(p,q) &= \mathcal{D}_0(p)\delta(p-q) + {\hat V}(p,q) \\
\mathcal{D}_0 &= {D}_0\frac{1}{1+\alpha^{-1} {D}_0}\,,
\end{align}
where $\mathcal{D}_0$ is obtained by integration of the blocked UV
action for free-fields if we take $D_0=f^{-2}K_0^{-1}\cancel{p}$.   It 
satisfies the GW relation for free fields, 
$\{\gamma_5, \mathcal{D}_0\} = 2 \alpha^{-1} \mathcal{D}_0 \gamma_5
\mathcal{D}_0$. 
Then, using \bref{D-relation} for two Dirac operators, $\mathcal{D}$ and
$\mathcal{D}'$, 
we can express $\hat V$ in terms of $\Theta$ as
\begin{align}
{\hat V}(p,q) = \eta(p)\bigg(\Theta\frac{1}{1+\alpha^{-1}\eta
 \Theta}\bigg)(p,q)\eta(q)\,.
\label{V1}
\end{align}
It leads to the Dirac operator 
\begin{align}
\mathcal{D}(p,q) = \mathcal{D}_0(p)\delta(p-q) 
+ \eta(p)\bigg(\Theta \frac{1}{1+\alpha^{-1}\eta
 \Theta}\bigg)(p,q)\eta(q)\,.
\label{Dirac5}
\end{align}
\bref{Dirac5} should be compared with
\bref{Dirac-op1}, and are recognized to belong to our most general set
of the solutions to the WT identities. The discussion given here
motivates to take \bref{defTheta}.  We also notice that the field
redefinition \bref{field-red.2} does make sense in continuum theory, but
does not make sense on the lattice.  For free-field theory, it
becomes $\Psi'(p) =\eta(p) \Psi(p) $. On the lattice, $\eta(p)$ vanishes
in the region of momentum where the species doublers appear. 

We conclude that the standard chiral symmetry is realized even when 
symmetry-breaking blocking is performed for the Dirac fields. However,
the their interactions with the scalar fields inevitably become
non-linear.

\section{Summary and outlook}

In this paper we have given the WT identities for chiral symmetry
realized in the presence of a symmetry-breaking blocking for the Dirac
fields within the ERG framework. The WT identities have been obtained
using a functional method where the Wilson action is expressed as
functional integral over the blocked UV action.  We have solved the WT
identities as purely algebraic equations, assuming the Wilson action is
bilinear in the Dirac fields. The assumption makes it possible to divide
the WT identities into two sets of equations: One corresponds to a generalized
GW relation for the fermionic sector. The other is to fix counter
actions consisting of the scalar fields which are needed to cancel the
contribution arising from the Jacobian factor associated with the
non-linear chiral transformation on the Dirac fields. Here, 
chiral symmetry is realized in such a way that the change of the Wilson
action under non-linear symmetry transformations is canceled with the
change of the functional measure in the WT identities. Our family of
solutions contains some special solutions: One is the continuum analog of the
solution discussed in the lattice theory.  The other one is a kind of
perfect action obtained by integrating the blocked UV action over the
Dirac fields and identifying IR scalars with UV ones. 

In addition to the non-linear realization of symmetry, we have also
discussed linear realization of chiral symmetry. This is achieved by a
field redefinition for the fermionic variables. The chiral
transformations for new Dirac fields are given with an operator
extension of $\gamma_5$ or even $\gamma_5$ itself.  Accordingly, even if
we have started with symmetry-breaking regularization, it is possible
for us to end up with the standard representation of chiral symmetry. The
whole complexity appears in the Yukawa couplings where the Dirac fields
have non-polynomial interactions with scalar fields.   

The WT identities considered in this paper can be lifted to the quantum
master equation (QME) in the Batalin-Vilkovisky antifield formalism\cite{BV}. The
reduction of the WT identities with the Jacobian contributions to those
without them corresponds to reduction of the QME to the classical master
equation (CME) via a canonical transformation in the space of fields and
antifields. It will be shown that the reduction of the QME to the CME
happens in chiral symmetry considered here.
 
Since our method given here is expected to apply to other linear
symmetries such as supersymmetry and abelian gauge symmetry,  
it is interesting to discuss their generalized GW relations.      

\vspace{5mm}

\noindent
{\bf Acknowledgments}

\noindent 
We thank K. Itoh and M. Satoh for collaboration at early stage, and also
one of the referees for suggesting us construction of most general solutions.
One of us (Y.I) is very grateful to J.M. Pawlowski and F. Bruckmann 
for enlightening discussion. He also
would like to thank the Institute of Theoretical Physics in Heidelberg
for hospitality.  This work is
supported in part by the Grants-in-Aid for Scientific Research
No.22540270 from the Japan Society for the Promotion of Science.

\appendix

\section{Field Redefinition}

We will show that 
\begin{align}
L(p,q) 
= 
\eta^{-1}(p)
\frac{1}{1-\alpha^{-1}{\cal V}'}(p,q)
\eta(q)
\,.
\label{defL1}
\end{align}
is a solution to 
\begin{align}
\delta L(p,q) 
&=  
i\epsilon
\bigg(
\hat{\gamma}_5(p)L(p,q)-L(p,q)\hat{\gamma}_5(q) 
- 2\int_k\alpha^{-1}(p)\eta^{-1}(p)
\gamma_5
{\cal V}'(p,k) \eta(k)
L(k,q) \bigg)  \, .
\label{deltaL1}
\end{align} 

Chiral transformation of $L(p,q)$ in \bref{defL1} is given by   
\begin{align}
\delta L(p,q) 
= 
- \int_{k,l}\eta^{-1}(p)
\frac{1}{1-\alpha^{-1}{\cal V}'}(p,l)\big(-\alpha^{-1}(l)\delta{\cal V}'(l,k)\big)
\frac{1}{1-\alpha^{-1}{\cal V}'}(k,q)\eta(q) \, .
\end{align}
Rewriting \bref{deltaTheta1} for $\delta{\cal V}'(p,q)$ as 
\begin{align}
\alpha^{-1}(l)\delta{\cal V}'(l,k) 
&= 
-2i\epsilon\alpha^{-1}(l)\bigg( 
\gamma_5{\cal V}'(l,k)-\int_q{\cal V}'(l,q)\gamma_5\alpha^{-1}(q){\cal V}'(q,k)\bigg) 
+ i\epsilon\alpha^{-1}(l)\big[\gamma_5\,,\,{\cal V}'(l,k)\big] \notag\\
&= 
-2i\epsilon\int_q\Big(\delta(l-q)-\alpha^{-1}(l){\cal V}'(l,q)\Big)
\gamma_5\alpha^{-1}(q){\cal V}'(q,k) 
+ i\epsilon\alpha^{-1}(l)\big[\gamma_5\,,\,{\cal V}'(l,k)\big]\, ,
\end{align}
we find
\begin{align}
\delta L(p,q) 
&= 
-2i\epsilon \int_{k}
\eta^{-1}(p)\gamma_5\alpha^{-1}(p){\cal V}'(p,k) 
\frac{1}{1-\alpha^{-1}{\cal V}'}(k,q)\eta(q) \notag\\
&\hspace{1truecm} 
+ i\epsilon \int_{k,l}
\eta^{-1}(p)
\frac{1}{1-\alpha^{-1}{\cal V}'}(p,l)\big[\gamma_5\,,\,\alpha^{-1}(l){\cal V}'(l,k)\big]
\frac{1}{1-\alpha^{-1}{\cal V}'}(k,q)\eta(q) \notag\\
& = -2i\epsilon\int_k\alpha^{-1}(p)\eta^{-1}(p) 
\gamma_5{\cal V}'(p,k)\eta(k) L(k,q) 
+ i\epsilon
\eta^{-1}(p)
\bigg[\gamma_5\,,\,\frac{1}{1-\alpha^{-1}{\cal V}'}(p,q)\bigg] 
\eta(q) \notag\\
&=i\epsilon\Big( 
\hat{\gamma}_5(p)L(p,q)-L(p,q)\hat{\gamma}_5(q) 
- 2\int_k\alpha^{-1}(p)\eta^{-1}(p)\gamma_5
{\cal V}'(p,k) \eta(k)L(k,q)\Big) \, , 
\end{align}
where we have used $\eta^{-1}(p)\gamma_5 = \hat\gamma_5\eta^{-1}(p)$. 


\section{The New Dirac Operator}

The new Dirac operator is given by
\begin{align}
{\cal D}'(p,q)  &=\int_{k} {\cal D}_{0}(p)\delta(p-k)\eta^{-1}(k) 
\frac{1}{1- \alpha^{-1} {\cal V}'}(k,q) \eta(q) 
+ \gamma_{5}\eta^{-1}(p) \gamma_{5}\int_{k}{\cal V}'(p,k)
\frac{1}{1- \alpha^{-1} {\cal V}'}(k,q)\eta(q) \notag\\
&= {\cal D}_{0}(p)\delta(p-q) + \Big[\eta^{-1}\alpha^{-1}{\cal D}_{0}
+  (\gamma_{5}\eta \gamma_{5})^{-1}\Big](p)\int_{k}{\cal V}'(p,k)
\frac{1}{1- \alpha^{-1} {\cal V}'}(k,q)\eta(q) \,.
\end{align} 
Using the GW relation $\{\gamma_{5}, {\cal D}_{0}\} = 2 \alpha^{-1} 
{\cal D}_{0} \gamma_{5}{\cal D}_{0}$, we find 
\begin{align}
\eta \gamma_{5} \eta \gamma_{5}= (1- \alpha^{-1}{\cal D}_{0})
[\gamma_{5}(1- \alpha^{-1}{\cal D}_{0})\gamma_{5}] 
= 1 - \frac{1}{2}\alpha^{-1} ({\cal D}_{0} + \gamma_{5} {\cal
 D}_{0}\gamma_{5})\, .
\end{align}
It leads to 
\begin{align}
& \eta^{-1}\alpha^{-1}{\cal D}_{0} + (\gamma_{5}\eta \gamma_{5})^{-1} = 
\biggl(\frac{1}{1 - \alpha^{-}{\cal D}_{0}}\alpha^{-1}{\cal D}_{0}
\biggr)+ \frac{1}{\gamma_{5}(1- \alpha^{-1}{\cal D}_{0})\gamma_{5}} \nn\\
& \quad = \biggl(\frac{1}{1 - \alpha^{-}{\cal D}_{0}}\biggr)
\biggl[\frac{1}{\gamma_{5}(1- \alpha^{-1}{\cal D}_{0})\gamma_{5}}\biggr]
\Big[\gamma_{5}(1 - \alpha^{-}{\cal D}_{0})
\gamma_{5}\alpha^{-1}{\cal D}_{0} + 1 - \alpha^{-1}{\cal D}_{0}  
\Big] = 1\, .
\end{align}
Therefore, 
\begin{align}
{\cal D}'(p,q)  
&= {\cal D}_{0}(p)\delta(p-q)  
+  \int_{k}{\cal V}'(p,k)
\Big(\frac{1}{1- \alpha^{-1} {\cal V}'}\Big)(k,q)\eta(q) \,.
\end{align} 

We next consider the definition of ${\cal V}'$ given in \bref{Theta'1}. 
It follows that
\begin{align}
\int_{k}{\cal V}'(p,k)(1 + B\Theta)(k,q) 
&= \frac{1}{2}\alpha(p)\gamma_{5} \{\gamma_{5}, B(p)\}
 \Theta(p,q) \nn\\
\int_{k}\Big(1 - \alpha^{-1} {\cal V}'\Big)(p,k) (1 + B\Theta)(k,q) 
&= \Big(1+  \frac{1}{2}\gamma_{5}[\gamma_{5}, B]\Theta\Big)(p,q)
\, .
\end{align} 
These relations give
\begin{align}
& \int_{k} {\cal V}'(p,k) \Big(\frac{1}{1 - \alpha^{-1} {\cal V}'}
\Big) (k,q) \equiv  \Big({\cal V}' \frac{1}{1 - \alpha^{-1} {\cal V}'}
\Big) (p,q) =\Big(\frac{1}{1 - {\cal V}' \alpha^{-1} }{\cal V}'
\Big)(p,q) \nn\\
&=\frac{1}{2}\int_{k}\alpha(p)\gamma_{5} \{\gamma_{5}, B(p)\}
\Theta(p,k)\biggl(\frac{1}{1+ \frac{1}{2}\gamma_{5}[\gamma_{5}, B]
\Theta}\biggr)(k,q) \nn\\
& =
\frac{1}{2}\alpha(p)\gamma_{5} \{\gamma_{5}, B(p)\}
\biggl(\Theta \frac{1}{1+ \frac{1}{2}\gamma_{5}[\gamma_{5}, B]
\Theta}\biggr)(p,q) \, .
\end{align} 

Since ${\cal V}'$ transforms as
\begin{align}
\delta {\cal V}' = -i \varepsilon \Big(
\gamma_{5} {\cal V}' + {\cal V}' \gamma_{5} -2 {\cal V}' \alpha^{-1}\gamma_{5}
{\cal V}' \Big) \, ,
\end{align} 
we have
\begin{align}
\delta \Big({\cal V}' \frac{1}{1 - \alpha^{-1} {\cal V}'}\Big)(p,q) =  
-i \varepsilon \bigg[
\gamma_{5} \Big({\cal V}' \frac{1}{1 - \alpha^{-1} {\cal V}'}\Big)(p,q) + 
\Big({\cal V}' \frac{1}{1 - \alpha^{-1} {\cal V}'}\Big)(p,q) \gamma_{5}
\bigg]\, .
\end{align} 
Using $\gamma_{5} \eta(q) = \eta \hat\gamma_{5}(q)$, we find 
\begin{align}
\delta {\cal V}''(p,q) = \delta \bigg[
\Big({\cal V}' \frac{1}{1 - \alpha^{-1} {\cal V}'}\Big)(p,q) \eta(q)
\bigg] =  -i \varepsilon \Big(\gamma_{5} {\cal V}''(p,q)
+  {\cal V}''(p,q) \hat\gamma_{5}(q) \Big) \,.
\end{align}


\end{document}